\documentclass[aps,prb,floatfix,twocolumn,superscriptaddress,showpacs,amsmath,amssymb]{revtex4}
\usepackage{graphicx}
\usepackage{epstopdf}
\usepackage{epsfig} 
\usepackage{color}
\allowdisplaybreaks

\newcommand{\be}{\begin{equation}}
\newcommand{\ee}{\end{equation}}
\newcommand{\bea}{\begin{eqnarray}}
\newcommand{\eea}{\end{eqnarray}}
\newcommand{\ba}{\begin{eqnarray*}}
\newcommand{\ea}{\end{eqnarray*}}
\newcommand{\dagga}{{\phantom{\dagger}}}

\newcommand{\bQ}{\mathbf{Q}}

\newcommand{\bk}{\mathbf{k}}

\newcommand{\dis}{\displaystyle}

\newcommand{\fract}[2]{\frac{\dis #1}{\dis #2}}
\newcommand{\Tr}{\mathrm{Tr}}
\newcommand{\eqn}[1]{(\ref{#1})}

\begin{document}

\title{Finite-temperature Gutzwiller approximation and the phase diagram of a toy-model for V$_2$O$_3$}
\author{Matteo Sandri} 
\affiliation{International School for
  Advanced Studies (SISSA), and CNR-IOM Democritos, Via Bonomea
  265, I-34136 Trieste, Italy} 
\author{Massimo Capone} 
\affiliation{International School for
  Advanced Studies (SISSA), and CNR-IOM Democritos, Via Bonomea
  265, I-34136 Trieste, Italy} 
\author{Michele Fabrizio} 
\affiliation{International School for
  Advanced Studies (SISSA), and CNR-IOM Democritos, Via Bonomea
  265, I-34136 Trieste, Italy}

\date{\today} 


\begin{abstract}
We exploit exact inequalities that refer to the entropy of a distribution to 
derive a simple variational principle at finite temperature for
trial density matrices of Gutzwiller and Jastrow type. We use the
result to extend at finite temperature the Gutzwiller approximation,
which we apply to study a two-orbital model that we believe captures some essential features of V$_2$O$_3$. 
We indeed find that the phase diagram of the model bears many similarities to that of real vanadium sesquioxide. In addition, we show that in a Bethe lattice,  where the finite temperature Gutzwiller approximation provides a rigorous upper bound of the actual free energy,  the results compare 
well with the exact phase diagram obtained by the dynamical mean field theory. 
\end{abstract}
\maketitle

\section{Introduction}
\label{uno}

A genuine Mott insulator, where 
the insulating character is due exclusively to charge localization, is a very useful concept, 
physically conceivable, but never realized in the ground state of known correlated materials. Indeed, no system can sustain at zero temperature the residual entropy that would be associated with all 
other electronic degrees of freedom different from charge. As a result,  
Mott localization is always accompanied at low temperature by other
phenomena that freeze 
those degrees of freedom, for instance magnetic ordering or structural
distortions, which effectively turn the Mott insulator into a  conventional band insulator.
By this we mean the possibility of reproducing low-temperature static and often also dynamic properties of a supposed Mott  
insulator by an independent-particle scheme, no matter how sophisticated it is.\cite{VO2-LDA} 
However, even though it provides satisfactory results, an independent-particle scheme, like Hartree-Fock or DFT within LDA and its extensions, 
has a drawback: it can describe only the simultaneous 
locking of charge and other degrees of freedom, like spin or lattice,  while in a Mott insulator the charge 
freezes at a much higher energy scale than any other degree of freedom. 

A tool that can reveal this hierarchy of energy scales typical of a
Mott insulator is the temperature, which unveils the
profound difference between the excited states of a Mott insulator and
those of its ``band-insulator'' counterpart\cite{MottSlater}.

The typical example are antiferromagnetic 
Mott insulators, e.g. La$_2$CuO$_4$ or Cr-doped V$_2$O$_3$, whose conducting properties remain insulating-like also above the N\'eel temperature, while any independent particle scheme would predict metallic behavior as soon as N\'eel order melts.  
It is therefore important to have at disposal theoretical tools able to deal with strong correlations and finite temperature. One that is 
currently adopted is dynamical mean field theory
(DMFT)\cite{Review_DMFT_96} in combination with local-density
approximation (LDA)\cite{LDA+DMFT-Review} or the GW
approximation\cite{GW+DMFT}. These combined methods are extremely
reliable for correlated materials, but they can become extremely
cumbersome and numerically demanding, especially if full consistency on
the density is required. 

More recently, there have been several attempts to combine efficiently and self-consistently LDA with the Gutzwiller variational 
approach within the Gutzwiller
approximation (GA),\cite{LDA+Gutzwiller-Deng,LDA+Gutzwiller-Giovanni,LDA+Gutzwiller-Nicola}
which is less accurate but much simpler and less demanding than DMFT. So far, these attempts have been restricted to ground state properties, hence to zero temperature. A finite temperature extension of the GA has actually been proposed in Ref. \onlinecite{finite-T-GW}, but only in the 
simplest case of the one-band Hubbard model, where it was shown to reproduce qualitatively well the exact DMFT phase diagram.\cite{finite-T-GW} However, in order to tackle more realistic models, it would be desirable to have at disposal a finite temperature generalization of the GA able to deal with generic multi-band models. 

This is actually the aim of the present work that is organized as follows. First, in section \ref{due} we derive a rigorous upper-bound estimate of the free-energy of a many-body Hamiltonian within the class of Gutzwiller- and Jastrow-like variational density matrices.      
Next, in section \ref{tre} we specialize to the case of Gutzwiller-like 
density matrices and introduce the Gutzwiller approximation at finite temperature. In section \ref{quattro}, we introduce a simple model for vanadium sesquioxide, V$_2$O$_3$, show how to implement in practice 
the Gutzwiller approximation at finite temperature, section \ref{quattro-A}, and discuss its results at 
zero and finite temperature, , section \ref{quattro-B} and  \ref{quattro-C}, respectively. 
In section \ref{cinque} 
we compare the phase diagram obtained by the Gutzwiller approximation on a Bethe lattice with the exact 
one obtained by DMFT. Finally, section \ref{sei} is devoted to concluding remarks. 

\section{Variational estimate of the free energy}
\label{due}
 
In this section we shall repeatedly use some known trace inequalities, 
for which we refer the reader to Ref. \onlinecite{Petz}. 
Let us consider an interacting many-body system described by the interacting Hamiltonian $\mathcal{H}$ at finite temperature $T>0$. It is known that 
the free-energy functional 
\be
F(X) = \Tr\big(X\,\mathcal{H}\big) + T\,\Tr\big(X\ln X\big),
\label{F(X)}
\ee
with the matrix $X>0$ and such that $\Tr\, X=1$, is minimized by 
the Boltzmann distribution function 
\be
X_\text{min} = \fract{\text{e}^{-\beta\mathcal{H}}}
{\Tr \, \text{e}^{-\beta\mathcal{H}}},\label{X-Boltzmann}
\ee
where $\beta=1/T$. Therefore, any variational ansatz for the density matrix $X$ provides an upper bound of the actual free energy
\be
F\equiv F\big(X_\text{min}\big) \leq F(X),  \quad \forall X> 0 \text{~with~} 
\Tr\,X=1.\label{variational-general}
\ee 
It is also known that, for any positive matrix $Y$, the entropy of the 
distribution $X$ satisfies the inequality\cite{Petz}
\bea
S(X) &=& -\Tr\big(X\ln X\big) \geq -\Tr\big(X\ln Y\big) \nonumber\\
&& 
- \Tr\Big(X\ln\big(X\, Y^{-1}\big)\Big)\equiv S_\text{var}(X,Y).\label{S(X)-ineq}
\eea
It then follows that, for any positive $Y$ and $X$ 
such that $\Tr \, X =1$,   
\bea
F\leq && \, \underset{X,Y}{\text{min}}\Big\{
\Tr\big(X\mathcal{H}\big) - T\,S_\text{var}(X,Y)
\Big\}.\label{variational-II}
\eea 
Eq. \eqn{variational-II} provides a variational principle for the free energy in terms of the distribution $X$ and the matrix $Y>0$. 
Let us assume the variational ansatz 
\be
X = \mathcal{P}\,\rho_*\,\mathcal{P}^\dagger,\label{ansatz}
\ee
where 
\be
\rho_* = \fract{\text{e}^{-\beta\mathcal{H}_*}}
{\Tr \, \text{e}^{-\beta\mathcal{H}_*}},\label{rho_*}
\ee 
is the Boltzmann distribution corresponding to a variational non-interacting Hamiltonian $\mathcal{H}_*$,
and $\mathcal{P}$ a many-body 
operator that we can parametrize as 
\be
\mathcal{P} = \mathcal{U}\,\sqrt{\mathcal{Q}},\label{def-P-vs-Q}
\ee
with unitary $\mathcal{U}$ and $\mathcal{Q}>0$. It follows that the entropy of the distribution $X$
\bea
S(X) &=& -\Tr\Big(X\,\ln X\Big) \nonumber\\
&=& -\Tr
\Big(\mathcal{Q}^{1/2}\,\rho_*\,\mathcal{Q}^{1/2}
\ln \big(\mathcal{Q}^{1/2}\,\rho_*\,\mathcal{Q}^{1/2}\big)\Big),
\eea   
is independent of the unitary operator $\mathcal{U}$. By means of 
Eq. \eqn{S(X)-ineq}, setting $Y=\mathcal{Q}$, we obtain
\bea
&& S_\text{var}(X,Y)  = -\Tr\Big( \mathcal{Q}^{1/2}\,\rho_*\,\mathcal{Q}^{1/2}
\,\ln \mathcal{Q}\Big)\nonumber\\
&& \qquad \qquad - \Tr\Big( \mathcal{Q}^{1/2}\,\rho_*\,\mathcal{Q}^{1/2}\,
\ln\big(\mathcal{Q}^{1/2}\,\rho_*\,\mathcal{Q}^{-1/2}\big)\Big)
\nonumber\\
&& \qquad = -\Tr\Big(\rho_*\,\mathcal{Q}
\,\ln \mathcal{Q}\Big) - \Tr\Big(\rho_*\,\mathcal{Q}\,
\ln\big(\rho_*\big)\Big).\label{entropy-final}
\eea
In conclusion, given the ansatz Eqs. \eqn{ansatz}-\eqn{def-P-vs-Q}, 
one can obtain an upper estimate of the actual free energy
\bea
F \leq && \, \text{min}\bigg\{ 
\Tr\Big(\rho_*\,\mathcal{P}^\dagger\,\mathcal{H}\,
\mathcal{P}\Big) + T\,\Tr\Big(\rho_*\,\mathcal{P}^\dagger\mathcal{P}
\,\ln \mathcal{P}^\dagger\mathcal{P}\Big)\nonumber\\
&& ~~~~~~~~+T\,\Tr\Big(\rho_*\,\mathcal{P}^\dagger\mathcal{P}\,
\ln\rho_*\Big)\bigg\},\label{final-variational}
\eea  
minimizing with respect to a non-interacting 
Hamiltonian $\mathcal{H}_*$ and a many-body operator $\mathcal{P}$. 
This minimization is feasible only for particular choices of 
$\mathcal{P}$. For instance, if $\mathcal{P}=1$, Eq. 
\eqn{final-variational} reduces to the well-known Hartree-Fock variational estimate of the free energy. Another possibility is that 
$\mathcal{P}$ is a two-body Jastrow factor, which can be handled by 
the variational Monte Carlo statistical approach.\cite{Sorella_SR} 
In the next section, we shall consider still another class of operators 
$\mathcal{P}$, which can be dealt with analytically in the limit of infinite coordination lattices.  

We conclude by noting that, since Eq. \eqn{final-variational} is based on the lower bound estimate 
$S_\text{var}(X,Y)$ of 
the entropy of the distribution $X$, Eq.  \eqn{entropy-final}, there is no guarantee that, within a particular class of $Y$, such estimate is positive at any temperature, as the true entropy should be. Therefore, it is more appropriate 
to state that 
\be
S(X) \geq \, \underset{Y>0}{\text{Max}}\Big\{S_\text{var}(X,Y),0\Big\}.
\ee
We further mention that  Eq. \eqn{S(X)-ineq} 
is actually the $p=1$ case of the more general inequality\cite{Petz}
\bea
S(X) &=& -\Tr\big(X\ln X\big) \geq -\Tr\big(X\ln Y\big) \nonumber\\
&& 
- \frac{1}{p}\,\Tr\Big(X\ln\big(X^{p}\, Y^{-p}\big)\Big),\label{entropy-ineq.}
\eea
which becomes an equivalence as $p\to 0$. We cannot exclude that exploiting Eq. \eqn{entropy-ineq.}
one could get a better but still manageable estimate of the entropy, though we did not succeed. 

\section{The Gutzwiller approximation at finite $T$}
\label{tre}

We assume a Hamiltonian 
\be
\mathcal{H} = \sum_{i,j}\sum_{ab}\,
\big(t^{ab}_{ij}\,c^\dagger_{ia}c^\dagga_{jb}+H.c.\big) 
+ \sum_i\, \mathcal{H}_i,\label{H}
\ee
defined on a lattice with coordination number $z$, and hopping parameters $t^{ab}_{ij}$ such that their 
contribution to the total energy is well behaved also in the limit $z\to\infty$. 
$\mathcal{H}_i$ includes on-site potential and 
interaction terms, while $c^\dagger_{ia}$ creates an electron at site $i$ with quantum number $a$, which 
can include spin and orbital.  
Given a variational density matrix of the form as in Eqs. \eqn{ansatz} and \eqn{rho_*}, i.e. 
\be
\rho_\text{G} = \mathcal{P}\,\rho_*\,\mathcal{P}^\dagger,\label{ansatz-G}
\ee
we take the operator $\mathcal{P}$ to be of the Gutzwiller type,\cite{Gutzwiller_1,Gutzwiller_2} i.e.
\be
\mathcal{P} = \prod_i\,\mathcal{P}_i,\label{P-G}
\ee
where $\mathcal{P}_i$ acts on the Hilbert space at site $i$. 
We shall impose that 
\bea
\Tr\Big(\rho_*\,\mathcal{P}_i^\dagger \mathcal{P}_i^\dagga\Big) &=& 1,\label{c-1}\\
\Tr\Big(\rho_*\,\mathcal{P}_i^\dagger \mathcal{P}_i^\dagga\,\mathcal{C}_i\Big) &=& 
\Tr\Big(\rho_*\,\mathcal{C}_i\Big),\label{c-2}
\eea
where $\mathcal{C}_i$ is any single-particle operator at site $i$. The above conditions replace 
those at zero temperature\cite{Gebhard,mio_dimer,Hvar} and 
allow to analytically compute averages over the 
distribution function $\rho_\text{G}$ in the limit of infinite coordination number, $z\to\infty$ (the proof is 
exactly the same as in Ref. \onlinecite{Hvar}, hence we shall not repeat it here). Specifically, 
when $z\to \infty$, the two conditions \eqn{c-1} and \eqn{c-2} imply that the distribution  $\rho_\text{G}$ 
has unit trace, and that all the formulas of Ref. \onlinecite{Hvar} hold 
with the only difference that, instead of averaging over a variational Slater determinant, one has to average 
over the variational non-interacting Boltzmann distribution $\rho_*$.  Following Ref. \onlinecite{Hvar}, 
we assume that $\rho_*$ identifies a local natural basis, with creation operators $d^\dagger_{i\alpha}$, 
related by a unitary transformation to the original operators $c^\dagger_{ia}$,  such that 
\be
\Tr\Big(\rho_*\, d^\dagger_{i\alpha}d^\dagga_{i\beta}\Big) = \delta_{\alpha\beta}\,n^{0}_{i\alpha},
\label{natural}
\ee 
where $n^{0}_{i\alpha}$ depends on the variational Hamiltonian $\mathcal{H}_*$ and on the temperature.
We also introduce local Fock states in the natural representation 
\be
\mid i;\{n\}\rangle = \prod_\alpha \big(d^\dagger_{i\alpha}\big)^{n_\alpha}\mid 0\rangle,
\ee
whose local probability distribution is diagonal, 
\ba
&&\Tr\Big(\rho_*\mid i;\{n\}\rangle\langle i;\{m\}\mid\Big) = \delta_{\{n\}\{m\}}\,
P^0_{i;\{n\}}\\
&&~~~~~~~~= \delta_{\{n\}\{m\}}\,\prod_\alpha\, \big(n^0_{i\alpha}\big)^{n_\alpha}\,
\big(1-n^0_{i\alpha}\big)^{1-n_\alpha}.
\ea
We then parametrize the operator $\mathcal{P}_i$ as
\be
\mathcal{P}_i = \sum_{\{n\},\Gamma}\,
\Phi_{i;\Gamma\{n\}}\,\Big(P^0_{i;\{n\}}\Big)^{-1/2}\mid i;\Gamma\rangle\langle i;\{n\}\mid,
\ee
where $\Phi_{i;\Gamma\{n\}}$ are the components of a temperature-dependent 
variational matrix $\Phi_i$, and $\mid i;\Gamma\rangle$ are local basis states in the original representation. 
In terms of $\Phi_i$ the Eqs. \eqn{c-1} and \eqn{c-2} read
\bea
\Tr\Big(\Phi_i^\dagger\Phi_i^\dagga\Big) &=& 1,\label{d-1}\\
\Tr\Big(\Phi_i^\dagger\Phi_i^\dagga\,d^\dagger_{i\alpha}
d^\dagga_{i\alpha}\Big) &=& n^{0}_{i\alpha},\label{d-2}
\eea
all other bilinear operators in \eqn{d-2} having null average. If we discard 
for simplicity the possibility of superconductivity, it follows that\cite{Hvar}
\bea
\Tr\Big(\rho_\text{G}\,\mathcal{H}\Big) &=& \sum_{i,j}\sum_{\alpha\beta}\,\Tr\Big(\rho_*\,
\big(t^{\alpha\beta}_{*\,ij}\,d^\dagger_{i\alpha}d^\dagga_{j\beta}+H.c.\big) \Big)\nonumber\\
&& + \sum_i\, \Tr\Big(\Phi_i^\dagger\,\mathcal{H}_i\,\Phi_i^\dagga\Big),\label{aver-H}
\eea
where 
\be
t^{\alpha\beta}_{*\,ij} = \sum_{ab}\, R^\dagger_{i\, \alpha a}\,t^{ab}_{ij}\,R^\dagga_{j\,b\beta},
\label{t_*}
\ee
with the renormalization factors  
\be
R^\dagger_{i\,\alpha a} = \big(n^0_{i\alpha}\left(1-n^0_{i\alpha}\right)\big)^{-1/2}\, 
\Tr\Big(\Phi_i^\dagger\,c^\dagger_{ia}\,\Phi_i^\dagga\,d^\dagga_{i\alpha}\Big).\label{R}
\ee
In other words, the average over $\rho_\text{G}$ of the Hamiltonian \eqn{H} is equal to the average   
over the uncorrelated distribution $\rho_*$ of a renormalized 
hopping Hamiltonian plus the sum of local terms that 
depend only on the variational matrices $\Phi_i$. 

We next need to evaluate the entropy.  We note that, in the $z\to\infty$ limit, and for any, even non-local,  
single-particle operator $\mathcal{C}$
\[
\Tr\Big(\rho_*\, \mathcal{P}^\dagger\mathcal{P}\,\mathcal{C}\Big)
 = \Tr\Big(\rho_*\, \mathcal{C}\Big).
\]
 Since it also holds that 
 \ba
 &&\Tr\Big(\rho_*\, \mathcal{P}^\dagger\mathcal{P}\,\ln \mathcal{P}^\dagger\mathcal{P}\Big)
 = \sum_i\, \Tr\Big(\rho_*\, \mathcal{P}_i^\dagger\mathcal{P}_i^\dagga\,
 \ln \mathcal{P}_i^\dagger\mathcal{P}_i^\dagga\Big)\\
 && = \sum_i\, \Tr\bigg[\Phi_i^\dagger \Phi_i^\dagga\, 
 \ln  \Big(\left(P_i^0\right)^{-1}\,\Phi_i^\dagger \Phi_i^\dagga\Big)\bigg], 
 \ea
it follows that Eq. \eqn{entropy-final} reads, in the $z\to\infty$ limit, 
\bea
S_\text{var}\big(\rho_*,\Phi^\dagger\Phi\big) &=&\,  S\big(\rho_*\big) \nonumber\\
&& \, - \sum_i\, \Tr\bigg[\Phi_i^\dagger \Phi_i^\dagga\, 
 \ln  \Big(\left(P_i^0\right)^{-1}\,\Phi_i^\dagger \Phi_i^\dagga\Big)\bigg]\nonumber\\
 && = S\big(\rho_*\big) + \sum_i\, S\Big(\Phi_i^\dagger \Phi_i^\dagga\big|| P_i^0\Big),
\eea
where $S\Big(\Phi_i^\dagger \Phi_i^\dagga\big|| P_i^0\Big)$ is the relative entropy between the distribution 
$\Phi_i^\dagger \Phi_i^\dagga$ and the uncorrelated local distribution $P_i^0$. In conclusion, the free energy can be upper estimated through 
\bea
F \leq && \, \text{min}\bigg\{  \sum_{i,j}\sum_{\alpha\beta}\,\Tr\Big(\rho_*\,
\big(t^{\alpha\beta}_{*\,ij}\,d^\dagger_{i\alpha}d^\dagga_{j\beta}+H.c.\big) \Big)
\nonumber\\
&&  ~~~~~~~~
+ \sum_i\, \Tr\Big(\Phi_i^\dagger\,\mathcal{H}_i\,\Phi_i^\dagga\Big) \nonumber\\
&& \qquad \quad - T\,\text{Max}\Big(S_\text{var}\big(\rho_*,\Phi^\dagger\Phi\big) ,0\Big)
\bigg\},\label{F-GW}
\eea
hence one just needs to minimize the r.h.s. supplemented by the constraints \eqn{d-1} and \eqn{d-2}. 
A possible route is to regard $n^0_{i\alpha}$ in Eqs. \eqn{natural} and \eqn{d-2} 
as independent minimization parameters, and introduce two Lagrange multipliers terms
\[
\Tr\Big(\rho_*\, \mathcal{V}\Big) - \sum_i\sum_{\alpha\beta}\, \mu_{i\,\alpha\beta}
\bigg[\Tr\Big(\Phi^\dagger_i
\Phi^\dagga_i\,d^\dagger_{i\alpha}d^\dagga_{i\beta}\Big) - \delta_{\alpha\beta}\,n^0_{i\alpha}\bigg],
\]
where the non-interacting potential $\mathcal{V}$ enforces Eq. \eqn{natural}, while  $\mu_{i\,\alpha\beta}$ enforce Eq. \eqn{d-2}. 

When $S_\text{var}\big(\rho_*,\Phi^\dagger\Phi\big) >0$, 
the saddle point with respect to the uncorrelated distribution $\rho_*$, see Eq. \eqn{rho_*}, corresponds 
to identifying 
\be
\mathcal{H}_* = \sum_{i,j}\sum_{\alpha\beta}\,
\big(t^{\alpha\beta}_{*\,ij}\,d^\dagger_{i\alpha}d^\dagga_{j\beta}+H.c.\big) + \mathcal{V},
\label{H_*}
\ee
so that, once $\mathcal{V}$ is chosen so as to satisfy Eq. \eqn{natural}, Eq. \eqn{F-GW} reads
\bea
F &&\leq  \, \text{min}\Bigg\{ F_*\big[\Phi,n^0\big] \nonumber\\
&& + \sum_i\, \Tr\Big(\Phi_i^\dagger\,\mathcal{H}_i\,\Phi_i^\dagga\Big) + S\Big(\Phi_i^\dagger \Phi_i^\dagga\big|| P_i^0\Big)\nonumber\\
&& - \sum_i\sum_{\alpha\beta}\, \mu_{i\,\alpha\beta}
\bigg[\Tr\Big(\Phi^\dagger_i
\Phi^\dagga_i\,d^\dagger_{i\alpha}d^\dagga_{i\beta}\Big) - \delta_{\alpha\beta}\,n^0_{i\alpha}\bigg]
\Bigg\}\nonumber\\
&& \equiv \underset{\Phi,n^0,\mu}{\text{min}}\Big\{ F\big[\Phi,n^0,\mu\big]\Big\},\label{final-F-GW}
\eea
where $F_*$ is the free energy of non-interacting electrons described by the Hamiltonian 
$\mathcal{H}_*$ in \eqn{H_*} that depends on the variational matrices 
$\Phi_i$ and on the parameters $n^0_{i\alpha}$ through  the constraint \eqn{natural} and 
the Eqs. \eqn{t_*} and \eqn{R}. 

However, when $T\to 0$, $S_\text{var}\big(\rho_*,\Phi^\dagger\Phi\big)$ becomes very small and it may happen that the saddle point of Eq.  \eqn{F-GW} is only a relative minimum, while the actual minimum is obtained 
by $\rho_*$ being the projection onto the ground state of the Hamiltonian $\mathcal{H}_*$ 
in Eq. \eqn{H_*}. In this case, as well as when $S_\text{var}\big(\rho_*,\Phi^\dagger\Phi\big) \leq 0$, the variational estimate of the free energy coincides with that 
of the ground state energy, evidently a drawback of the entropy bound that we use. 
In our experience, this problem may arise only at very low temperature, in which case, although the free energy estimate is continuous, its derivative may not be so. This is evidently not a signal of a genuine first order transition but only a flaw of the method. However, since the temperature when this occurs is extremely low, the entropy contribution is nonetheless negligible, so is the error in the free energy.

Minimization of $F\big[\Phi,n^0,\mu\big]$ therefore provides an upper bound to 
the actual free energy in lattices with infinite coordination number $z\to\infty$. Seemingly to what it is 
done at zero temperature, one can keep using the same free-energy functional also when the coordination number is finite, which can be regarded as the finite temperature extension of the Gutzwiller 
approximation.\cite{Gutzwiller_1,Gutzwiller_2,Gebhard} We mention that, in the simple case of a 
one-band Hubbard model, the free energy functional $F\big[\Phi,n^0,\mu\big]$ coincides with the expression 
derived by different arguments in Ref. \onlinecite{finite-T-GW}. 

In the next section we apply the outlined method to a simple two-band Hubbard model and test its results with exact ones obtained by DMFT. 

\section{A toy-model for V$_2$O$_3$}
\label{quattro}

The model we are going to analyze is inspired by the physics of 
V$_2$O$_3$. In this compound, the V$^{2+}$ ions have two valence electrons occupying the conduction bands that originate mainly from the $t_{2g}$ atomic $d$-orbitals. At high temperatures, V$_2$O$_3$ is 
a paramagnetic metal but, upon substituting V with Cr it can turn into a paramagnetic 
insulator.\cite{McWhan&Remeika&Rice} The transition is first order and
ends into a second-order critical point. 
At low temperature, V$_2$O$_3$ is instead an antiferromagnetic insulator. The N\'eel transition occurs at 
$T_N\simeq 170$ K and is accompanied by a martensitic transformation from the high-temperature corundum structure to the low-temperature monoclinic one.\cite{McWhan&Remeika} As a result, the 
magnetic ordering is not a simple G-type, as it could well be in a bipartite lattice, but, in the 
honeycomb-lattice basal plane, two bonds are antiferromagnetic and one is ferromagnetic.\cite{Moon} 
There is wide consensus that the magnetic moment is formed by  a spin $S=1$\cite{Park&Sawatzky} 
but it is also contributed by angular momentum,\cite{Paolasini} signaling a non-negligible spin-orbit coupling. Even though a reliable description of the antiferromagnetic transition requires including 
electron-lattice and spin-orbit couplings, the main features of the phase diagram can be likely explained 
ignoring those additional complications. The trigonal field of the corundum structure splits the $t_{2g}$ orbitals into a lower $e^\pi_g$ doublet and a higher $a_{1g}$ singlet. It is therefore tempting to conclude that the low-temperature insulator describes the two electrons in the $e^\pi_g$ doublet that, because of Hund's rules, are coupled into a spin-triplet. This conclusion is probably not far from reality. Indeed, 
although the bare value of the crystal field splitting is too small in comparison with the bare conduction bandwidth,\cite{Mattheiss} strong enough electronic correlations may reverse the situation and stabilize 
the insulating phase.\cite{Manini&Santoro} This scenario has been actually advocated to explain the phase diagram of V$_2$O$_3$ on the basis of a DMFT-LDA calculation in Ref. \onlinecite{Georges&Andersen}, and seems supported by some experimental evidences.\cite{Basov-V2O3} Indeed, DMFT-LDA results 
have shown that the effective crystal field splitting $\Delta_\text{eff}$ between $e^\pi_g$ and $a_{1g}$ orbitals is enhanced by correlations from its bare value $\Delta\simeq 0.27$ eV to one four times larger, which increases as the strength of the electron repulsion.\cite{Georges&Andersen} In addition, 
$\Delta_\text{eff}$ has been found to increase upon lowering temperature $T$, though only 
slightly,\cite{Georges&Andersen} but, more importantly, it has been observed that the magnetic 
susceptibility of the $e^\pi_g$ increases substantially with lowering $T$, while that of the $a_{1g}$ stays 
constants\cite{Georges&Andersen} or even diminishes,\cite{Grieger-V2O3} 
precursor signals of a magnetic instability that involves only $e^\pi_g$ orbitals.

However, all calculations so far 
have not been pushed down to the N\'eel transition temperature to really uncover the proposed mechanism of a gradual depopulation of the $a_{1g}$-derived band and concomitant magnetic polarization of the 
$e^\pi_g$-derived ones. Here, we would like to address this issue by exploiting the finite temperature 
technique described in section \ref{due} on a simplified  model that
we believe captures the essential physics. Instead of considering
three $t_{2g}$ orbitals split into two plus one and occupied on
average by two electrons, we shall  consider only two split orbitals
occupied on average by one electron (quarter filling). In this way we 
miss the important role of Coulomb exchange, which forces the two electrons on the $e^\pi_g$ doublet to lock into a spin triplet state and might bring about relevant incoherence effects,\cite{Luca-Janus} but the gross features of the phase diagram, in particular the interplay between temperature, crystal field splitting, correlations and magnetism, should be maintained.  

Specifically, we shall study the two-band Hamiltonian on a square lattice 
\begin{align}
&&\mathcal{H} = \sum_{a=1}^2\sum_{\bk\sigma}\, 
\epsilon_{\bk}\, c^\dagger_{a\bk\sigma}c^\dagga_{a\bk\sigma} 
+ \sum_{\bk\sigma}\,\gamma_\bk\, \big(c^\dagger_{1\bk\sigma}c^\dagga_{2\bk\sigma}+H.c.\big)
\nonumber\\
&& + \sum_i\bigg[
-\Delta\,\big(n_{1i}-n_{2i}\big) + \frac{U}{2}\,\big(n_{1i}+n_{2i}\big)^2\bigg],\label{Ham}
\end{align}
where $a=1,2$ labels the two orbitals, $\epsilon_\bk = -2t\big(\cos k_x + \cos k_y\big)$ is 
the standard nearest neighbor tight-binding energy, $U$ parametrizes the on-site repulsion and 
$\Delta>0$ the crystal field splitting. We include an inter-orbital hopping 
$\gamma_\bk = -4t'\,\sin k_x\sin k_y$ with a symmetry such that the 
local single-particle density matrix remains diagonal in the orbital indices 1 and 2, thus mimicking 
the $a_{1g}$-$e^\pi_g$ hybridization in the corundum phase of V$_2$O$_3$.\cite{Georges&Andersen} 
We shall further assume a density corresponding to one electron per site. 

In spite of its simplicity, the model in Eq. \eqn{Ham} 
reproduces qualitatively the actual behavior of  
V$_2$O$_3$. If $\Delta\simeq t'\ll t$, which we shall consider hereafter, the model describes a two-band metal for small $U$. However, for very large $U$, we do expect a Mott insulating phase with the electrons 
localized mostly on the lowest orbital and antiferromagnetically ordered. Therefore a strong repulsion $U$ 
can turn the two band metal into a single-band antiferromagnetic insulator, the two-band analogue of 
what is predicted in V$_2$O$_3$. The question we would like to 
address here is the behavior at finite temperature. 

We first observe that the enhancement of the 
effective crystal field $\Delta_\text{eff}$ caused by $U$, which eventually leads to antiferromagnetism  
once the highest band is emptied, can be described also within Hartree-Fock. Indeed, if we neglect magnetism and assume the variational mean-field ansatz  
\[
\langle n_{1i}\rangle = \frac{1}{2} + \delta n, \quad \langle n_{2i}\rangle = \frac{1}{2} - \delta n,
\]
then the Hartree-Fock energies of the orbitals are 
$\epsilon_1 = - \Delta + U\,(3-2\delta n)/2$, and $\epsilon_2 = \Delta + U\,(3+2\delta n)/2$, 
so that the effective crystal-field splitting is, within mean-field, 
$\Delta_\text{eff} = \Delta + U\,\delta n>\Delta$. As $U$ increases, $\Delta_\text{eff}$ grows hence the highest band depopulates until it becomes completely empty. Beyond this point, only the lowest band remains occupied, specifically half-filled, which can lead to a Stoner-like antiferromagnetic instability, hence to an insulating state.  In other words, as we mentioned in the 
Introduction, an independent particle picture, like Hartree-Fock, is indeed able to explain the occurrence of 
an antiferromagnetic insulating state at low temperature. However, no matter how large $U$ is, 
Hartree-Fock will predict this insulating phase to turn metallic above the N\'eel temperature $T_N$. On the contrary, we expect that, for $T> T_N$ but $U$ large enough, 
the phase should still be insulating, though paramagnetic. 

\subsection{Finite-$T$ Gutzwiller approximation at work}
\label{quattro-A}

We can improve the Hartree-Fock description at finite temperature by the Gutzwiller variational approach 
of section \ref{due}. By our choice, even though the inter-orbital hybridization $t'$ is finite, hence the two orbital can mutually exchange electrons, still the local density matrix is diagonal by symmetry. In other 
words, the natural basis, see Eq. \eqn{natural}, coincides with the original one; a great simplification in the calculations.  Since we will search for simple two-sublattice N\'eel order, we can set, for any site $i$ belonging to sublattice $A$, $\Phi_i = \Phi_A \equiv \Phi$ and $n^0_{i a\sigma} = n^0_{A a\sigma}\equiv n^0_{a\sigma}$ (see Eq. \eqn{natural}), such 
that $\sum_{a\sigma} n^0_{a\sigma} = 1$.  The variational 
matrix $\Phi$ is defined in the local Fock space and is only invariant under spin-rotations around the magnetization axis, which we choose as
the $z$ axis. It follows that, for any site $i$ belonging to the other sublattice $B$, 
$\Phi_i = \Phi_B = U^\dagger\,\Phi\,U$ with $U=\exp\big(i\pi S_y/2\Big)$ and $S_y$ 
the $y$-component of the total local spin,  while 
$n^0_{i a\sigma} = n^0_{B a\sigma}\equiv n^0_{a-\sigma}$.  Because of Eq. \eqn{d-2}, we 
must impose the constraint 
\be
\Tr\Big(\rho_* c^\dagger_{i a\sigma}c^\dagga_{i b\sigma'}\Big) = 
\Tr\Big(\Phi^\dagger\Phi \, c^\dagger_{ a\sigma}c^\dagga_{ b\sigma'}\Big) = 
\delta_{ab}\delta_{\sigma\sigma'}\, n^0_{a\sigma},\label{G-natural}
\ee
for $i\in A$, while, for $i\in B$, $\Phi\to U^\dagger\,\Phi\,U$ and $n^0_{a\sigma}\to n^0_{a-\sigma}$. 

Since natural and original bases coincide, the renormalization factors of Eq. \eqn{R} are diagonal 
and read, for $i\in A$, 
\be
R^*_{i\, a\sigma} = R^*_{a\sigma} = \fract{1}{\sqrt{n^0_{a\sigma}\left(1-n^0_{a\sigma}\right)}}
\Tr\Big(\Phi^\dagger\,c^\dagger_{a\sigma}\,\Phi^\dagga\,c^\dagga_{a\sigma}\Big),\label{R-2}
\ee
while, for $i\in B$, $R^*_{i\,a\sigma} = R^*_{a-\sigma}$. 
We find that, at the optimized values of the variational parameters, $R_{i\,a\sigma}$ are always real. Therefore, 
if we define $R_{a\sigma} \equiv 
R_a + \sigma\,S_a\in \Re e$, then the variational uncorrelated Hamiltonian $\mathcal{H}_*$, see Eq. \eqn{rho_*}, is
\bea
&&\mathcal{H}_* \big[\Phi,n^0\big] = \sum_{a=1}^2\sum_{\bk\sigma}\,\epsilon_{\bk a}\, 
c^\dagger_{a\bk\sigma}c^\dagga_{a\bk\sigma} \nonumber\\
&& \quad \quad+ 
\sum_{\bk\sigma}\,\gamma'_{\bk}\big(c^\dagger_{1\bk\sigma}c^\dagga_{2\bk\sigma} + H.c.\big)\nonumber\\
&& \quad\quad+ \sum_{\bk\sigma}\,\sigma\, \gamma"_{\bk}
\big(c^\dagger_{1\bk\sigma}c^\dagga_{2\bk+\bQ\sigma} + H.c.\big)\nonumber\\
&& \quad\quad+ \mu \sum_{a\bk\sigma} n_{a\bk\sigma} + 
\mu_\text{CF}\sum_{\bk\sigma}\big(n_{1\bk\sigma}-n_{2\bk\sigma}\big)\nonumber\\
&& \quad + \sum_{a=1}^2\sum_{\bk\sigma}\, 
\sigma\, h_{a}\big(c^\dagger_{a\bk\sigma}c^\dagga_{a\bk+\bQ\sigma}+ H.c.\big),\label{H_*-G}
\eea
where $\bQ=(\pi,\pi)$, the Lagrange multipliers $\mu$, $\mu_\text{CF}$, $h_1$ and $h_2$ enforce the constraints 
\eqn{G-natural}, and 
\ba
\epsilon_{a\bk} &=& \big(R_a^2 - S_a^2\big)\,\epsilon_\bk,\\
\gamma'_\bk &=& \big(R_1\,R_2+ S_1\,S_2\big)\,\gamma_\bk,\\
\gamma"_\bk &=& \big(R_1\,S_2+R_2\,S_1\big)\,\gamma_\bk.
\ea
It follows that, if 
\be
F_*\big[\Phi,n^0\big] = - \frac{T}{N}\ln \Tr\Big(\text{e}^{-\beta\,\mathcal{H}_*}\Big),
\ee
where $N$ in the number of sites, then we have to minimize
\bea
F\big[\lambda,\Phi,n^0\big] &=& F_*\big[\Phi,n^0\big]  +\fract{U}{2}\, \Tr \Big(\Phi^\dagger 
\big(n_1+n_2\big)^2\Phi\Big)\nonumber\\
&& - \Delta \Tr \Big(\Phi^\dagger \big(n_1-n_2\big)\Phi\Big)\nonumber\\
&& - \sum_{a\sigma}\, \lambda_{a\sigma}\,
\bigg[\Tr\Big(\Phi^\dagger\Phi\,n_{a\sigma}\Big) - n^0_{a\sigma}\bigg]\nonumber\\
&& - T\,S\big(\Phi^\dagger\Phi || P^0\big),\label{G-F}
\eea
with the constraint $\sum_{a\sigma} n^0_{a\sigma} = 1$. 

We find more convenient to minimize the variational free energy (\ref{G-F}) first with respect to all parameters except  
$n^0$,\cite{LDA+Gutzwiller-Nicola} thus obtaining the functional  
\be
F[n^0] = \underset{\lambda,\Phi}{\min}\, F\big[\lambda, \Phi,n^0 \big].
\ee
We calculate $F[n^0]$ in a two-step cyclic process; first we fix $\Phi$ and minimize $F_*[\Phi,n^0]$ with 
respect to the Lagrange multipliers in Eq. \eqn{H_*-G}.
Then, at fixed matrix elements $\langle c^\dagger_{ia\sigma}c^\dagga_{jb\sigma'}\rangle_{\rho_*}$, we minimize $F[\lambda,\Phi,n^0]$ with respect to $\Phi$ fulfilling the Gutzwiller constraints. This second non-linear constrained minimization is performed by the LANCELOT B routine of the GALAHAD library.\cite{Galahad} This two-step cycle is repeated until convergence. Finally a full minimization of $F[n^0]$ with respect to $n^0$ is performed.

\begin{figure}[!h]
\includegraphics[scale=0.32]{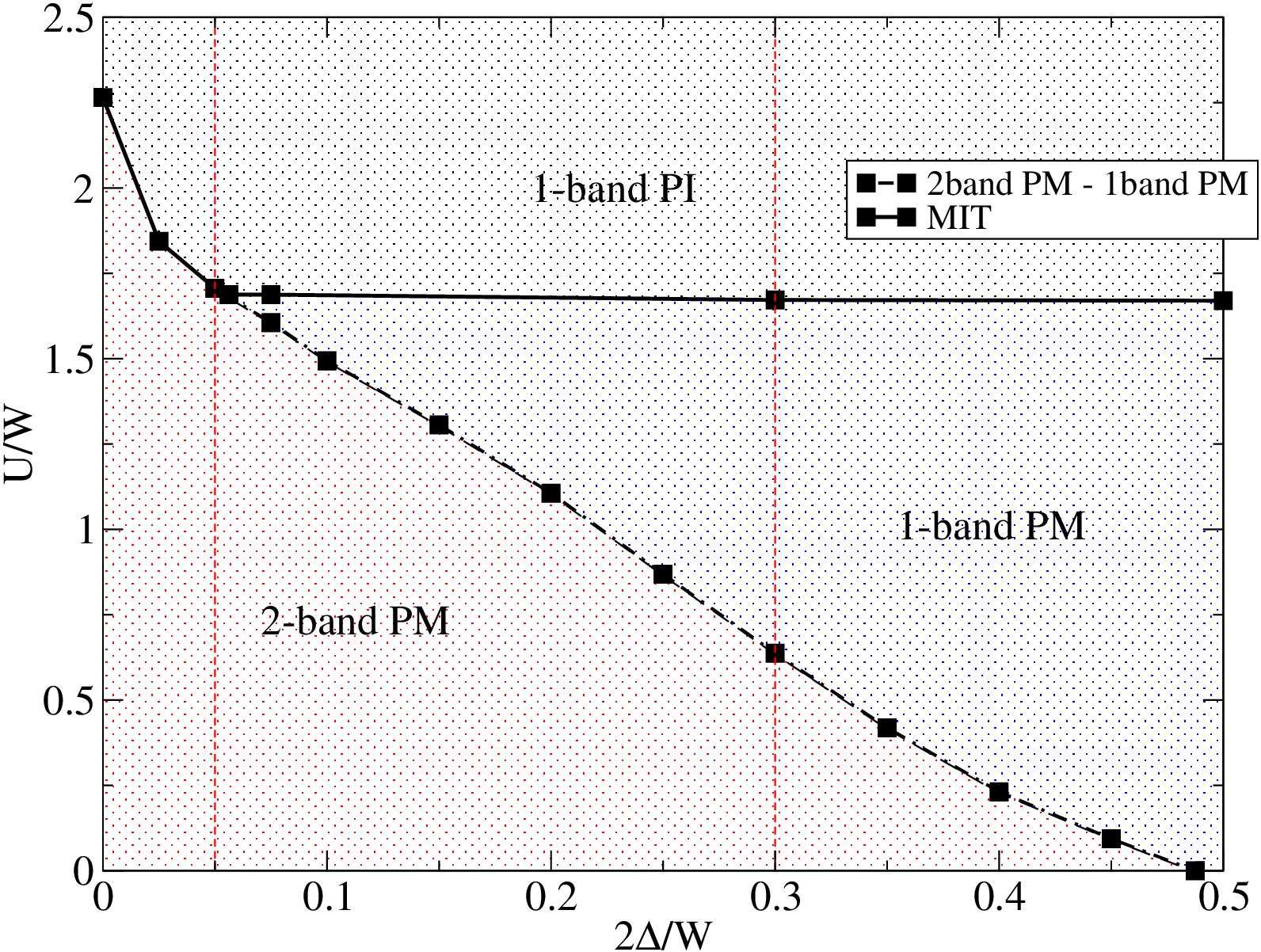}
\caption{(Color online) $T=0$ phase diagram for model in Eq. \eqn{Ham}. The solid line indicates the MIT transition; the dashed/dotted line separates the two-band paramagnetic metal (2-band PM) from the one-band metal 
(1-band PM). The vertical dashed red lines indicate the values of $\Delta$ which are used in 
Fig.\ref{fig:phaseT0_R}.}
\label{fig:phaseT0}
\end{figure}
\subsection{$T=0$ phase diagram}
\label{quattro-B}

The results that follow are obtained setting $t=1/8$ and the 
inter-orbital hybridization $t' = 0.3t$. The $t'=0$ bandwidth 
$W=8t=1$ hence sets the unit of energy.

First we consider the $T=0$ case of Eq. (\ref{G-F}), which corresponds to the usual Gutzwiller variational approach. 
In Fig.\ref{fig:phaseT0} we plot the zero temperature phase diagram 
in the paramagnetic sector as a function of $U$ and $\Delta$. Our results compare well with the DMFT phase diagram of 
Refs. \onlinecite{Manini&Santoro} and \onlinecite{Ferrero}.
In the limit $\Delta = 0$ the model undergoes a second order metal-to-insulator transition (MIT) at a critical value $U_c^{\Delta = 0} \simeq 2.27W$. In the opposite non interacting case, $U=0$, upon increasing $\Delta$ the system crosses a 
Lifshitz transition from a two-band to a one-band metal. 
We note that the majority ($>$) and minority ($<$) bands do not have a unique orbital character, therefore the band polarization $n_>-n_<$, which saturates to 1 at the two-band $\rightarrow$ one-band transition, is in general different from the orbital polarization $n_1 - n_2$.

\begin{figure}[th]
\includegraphics[scale=0.32]{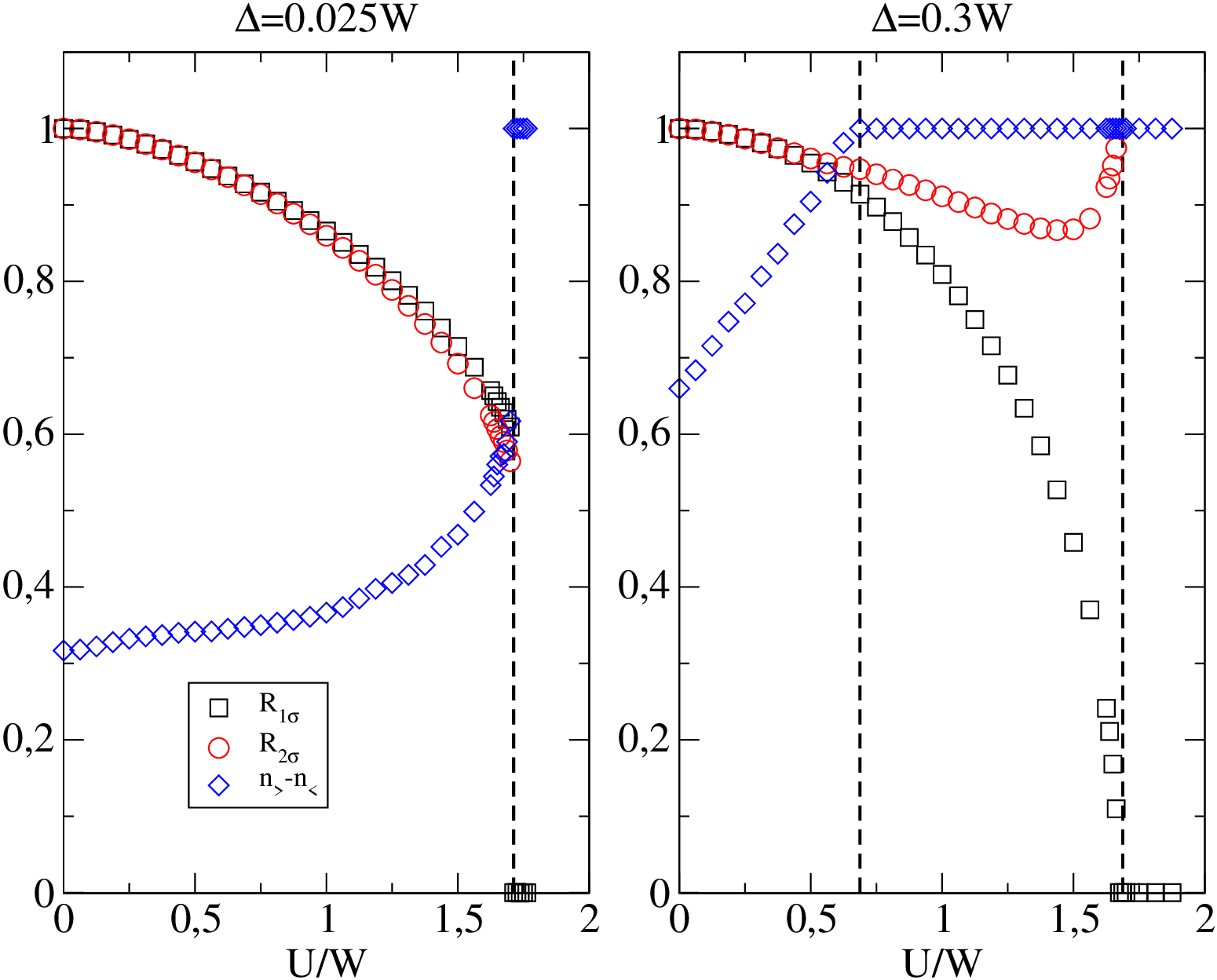}
\caption{(Color online) $n_{>} - n_{<}$ (blue diamonds), renormalization factors $R_{1}$ (black squares) and $R_{2}$ (red circles) as a function of $U$ for fixed $\Delta = 0.025W$ (left panel) and $\Delta = 0.3W$ (right panel). The vertical dashed lines indicate the two-band $\rightarrow$ one-band metal transition and the MIT.}
\label{fig:phaseT0_R}
\end{figure}

As anticipated, at finite $U$ a smaller crystal field spitting is required to induce the two-band $\rightarrow$ one-band transition, see Fig.\ref{fig:phaseT0}. Above this transition, the ground state is a one-band metal which eventually undergoes a second order MIT at a critical value $U_c\simeq 1.68W$. In Fig. \ref{fig:phaseT0_R} (right panel) we show the details of these two subsequent transitions for a value of $\Delta = 0.3W$. For $U \le 0.64W$, the two-band metal is stable but, increasing $U$, the minority band gradually empties and both renormalization factors, $R_1$ and $R_2$, decrease. At $U \simeq 0.64W$ the minority band completely depopulates and the leftover half-filled majority band is driven to the MIT at $U \simeq 1.68W$. 
Approaching the MIT, the renormalization factor $R_1$ of the lowest-energy orbital vanishes, while $R_2$ actually increases to one -- the almost empty orbital undresses from correlations.   
For smaller $\Delta$, the two-band $\rightarrow$ one-band transition becomes first order and approaches the one-band MIT point, ending in a multicritical point at $\Delta \simeq 0.028W$.
In Fig.\ref{fig:phaseT0_R} (left panel) we plot the behavior of $R_1$, 
$R_2$ and $\big(n_>-n_<\big)$ for $\Delta = 0.025W$; in this case the two renormalization factors are approximately equal and decrease monotonically with $U$.  At the transition, the majority orbital occupation suddenly increases and the corresponding renormalization factor vanishes. We mention that 
a Mott insulator with partial occupation of both orbitals can not be stabilized within the Gutzwiller approximation, while more reliable DMFT calculations show that 
such a phase does exist for very small $\Delta$.\cite{Ferrero}

\begin{figure}[t]
\includegraphics[scale=0.32]{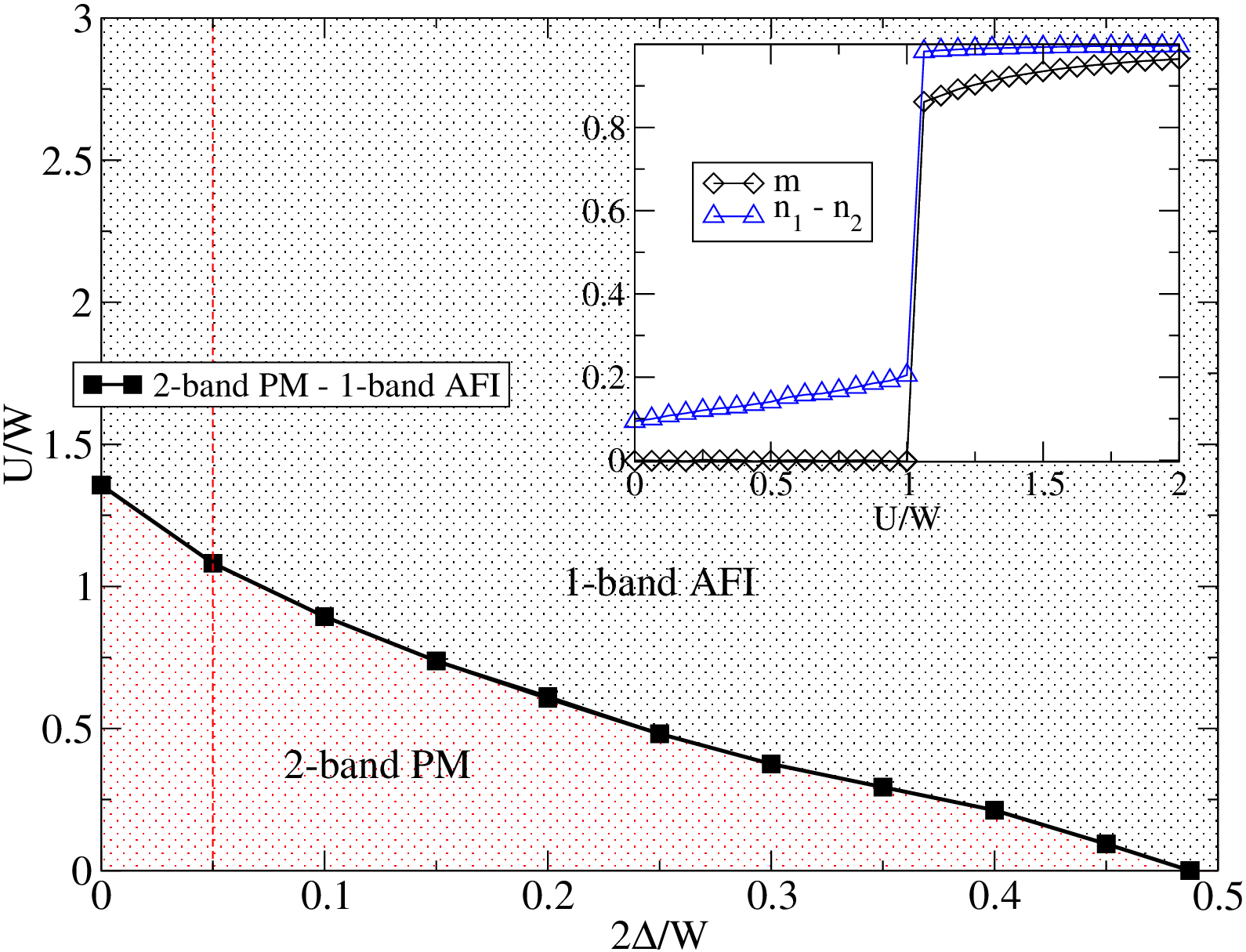}
\caption{(Color online) Zero temperature phase diagram allowing for magnetism. The black line separates the two-band paramagnetic metal (2-band PM) from the one-band antiferromagnetic insulator (1-band AFI). In the inset we plot the orbital polarization $n_{1} - n_{2}$ and the staggered magnetization for a fixed value of $\Delta = 0.025W$ (red dashed line).}
\label{fig:phaseT0_R_AF}
\end{figure}

If we allow for magnetism, the one-band phases, either metallic or Mott insulating, 
turn immediately into an antiferromagnetic insulator, see 
Fig.\ref{fig:phaseT0_R_AF}. 
The transition from the two-band paramagnetic metal (2-band PM) to the one-band antiferromagnetic insulator (1-band AFI) is first order and accompanied by a jump in the orbital polarization and in the staggered magnetization, see inset of 
Fig.\ref{fig:phaseT0_R_AF}. 

\subsection{$T\neq0$ phase diagram}
\label{quattro-C}

We have seen that at zero temperature the ground state is either a one-band antiferromagnetic insulator or a two-band paramagnetic metal. 
Therefore, as we anticipated, the $T=0$ Gutzwiller variational results are not dissimilar from the predictions of the Hartree-Fock approximation. Differences instead arise at finite 
temperature, where the Gutzwiller variational approach, as we are going to show, can describe melting  of the N\'eel order without metallization, unlike Hartree-Fock.  

\begin{figure}[!h]
\includegraphics[scale=0.32]{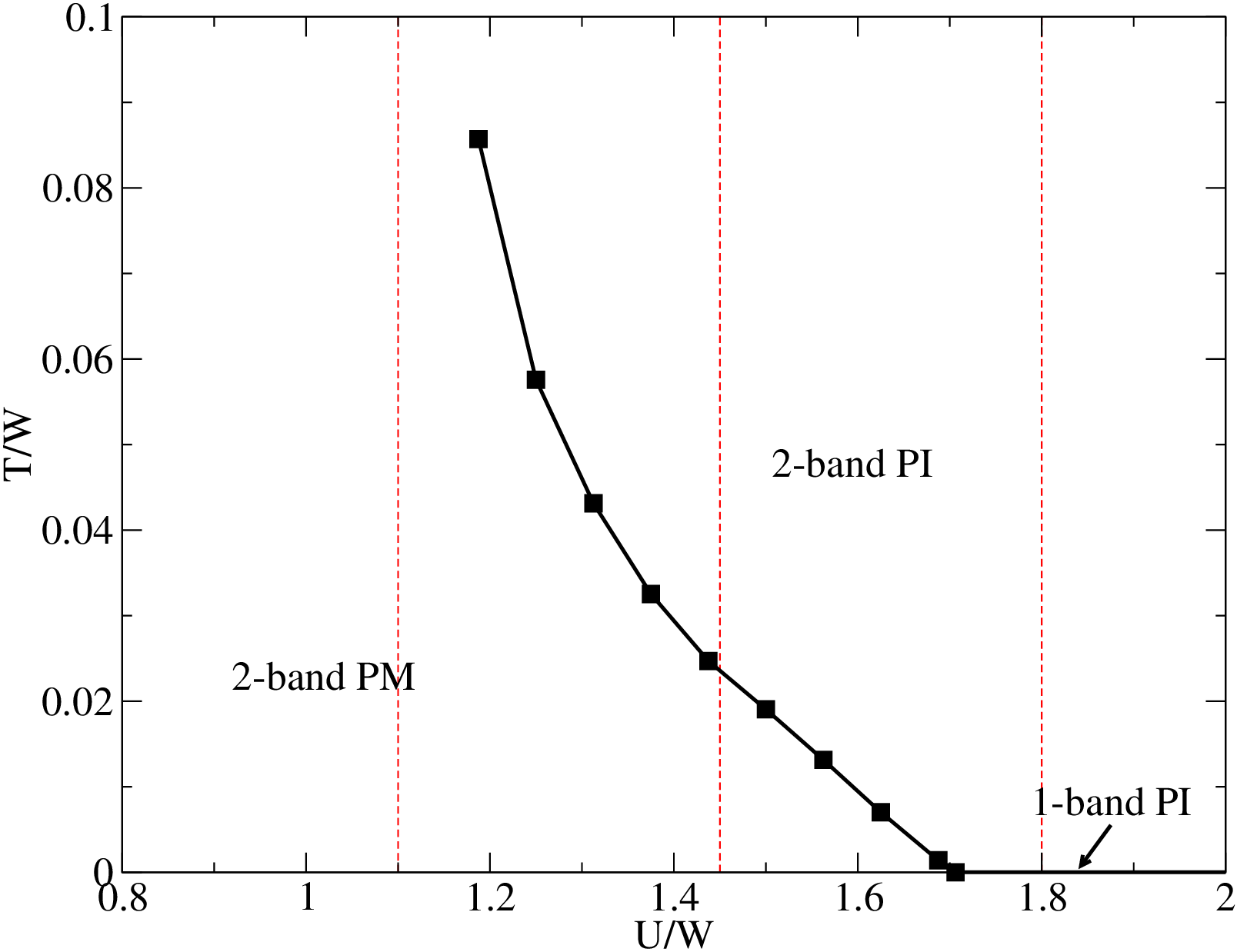}
\vspace{10pt}

\includegraphics[scale=0.32]{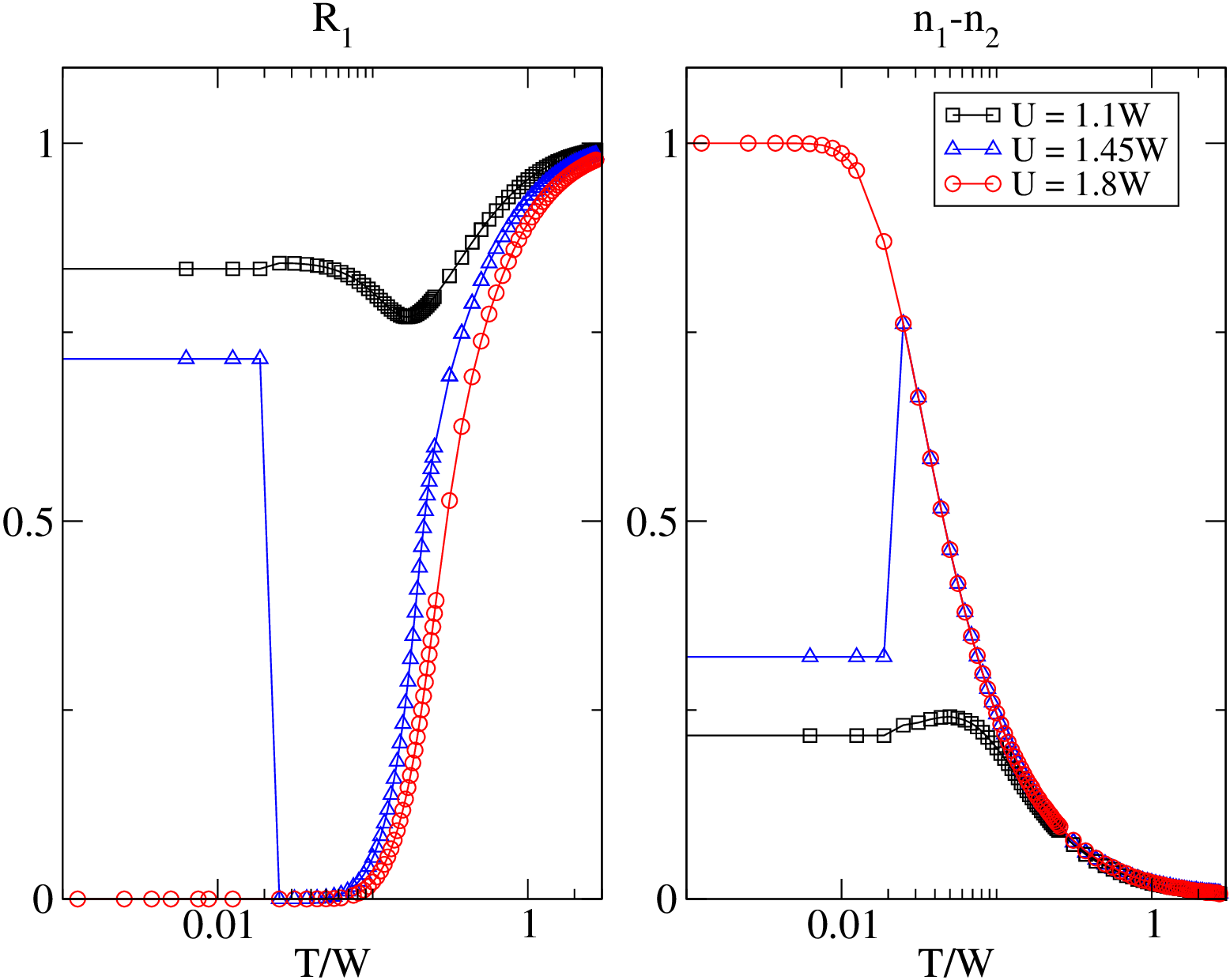}
\caption{(Color online) Upper panel: phase diagram in the paramagnetic domain. The black line separates the PM phase from the PI one. The red vertical lines indicate the values of $U$ plotted in the lower panel.\\
Lower panel: Temperature dependence of the quasiparticle renormalization factor for the majority orbital (left) and of the orbital polarization (right) for different values of $U$. }
\label{fig:phaseFT_PM}
\end{figure}

We begin as before by restricting the analysis to the paramagnetic sector and consider the case of $\Delta = 0.025W$. At zero temperature we found that the model is a 1-band PI for values of $U \ge 1.7W$, while 
 a 2-band PM below, Fig.\ref{fig:phaseT0}. At finite temperature, the entropic contribution may favor the paramagnetic insulating 
solution, like in the single band Hubbard model,\cite{Review_DMFT_96} thus leading to a finite $T$ metal-insulator transition.
This indeed occurs, as shown in Fig. \ref{fig:phaseFT_PM} where we plot the phase diagram as a function of $U$ and $T$ (upper panel) and the temperature dependence of the majority orbital $R_{1}$ and the orbital polarization (lower panels).
In the figure we observe that for values of $U\ge 1.7W$, increasing the temperature the orbital polarization decreases and the quasiparticle weight increase: the 1-band PI continuously evolves towards 
a 2-band PI.
Instead, for smaller values of $U$, the system is initially a two-band metal and undergoes a first order transition to an insulating state which is accompanied by an abrupt fall-down of the renormalization factors and increase of orbital polarization. As in Ref. \onlinecite{finite-T-GW}, we interpret the jump of the renormalization factor as the boundary of the PM-PI transition. Notice that, differently from the $T=0$ case, the orbital polarization does not saturate at the transition.
Finally, for values of $U$ smaller than $\sim 1.19W$, the quasiparticle weight and the orbital polarization evolve smoothly to the high temperature limit, displaying a dip that we interpret as the crossover regime. We estimate the end-point of the transition at $T\simeq0.09W$. 

We note, in the lower panel of  Fig. \ref{fig:phaseFT_PM} and for $U=1.1W$, the tiny  
discontinuity of $R_1$ and $n_1-n_2$ at $T\simeq 0.01W$, which is consequence of the aforementioned artificial discontinuity in the slope of the free energy caused by our not rigorous lower bound of the entropy. 

\begin{figure}[t]
\includegraphics[scale=0.32]{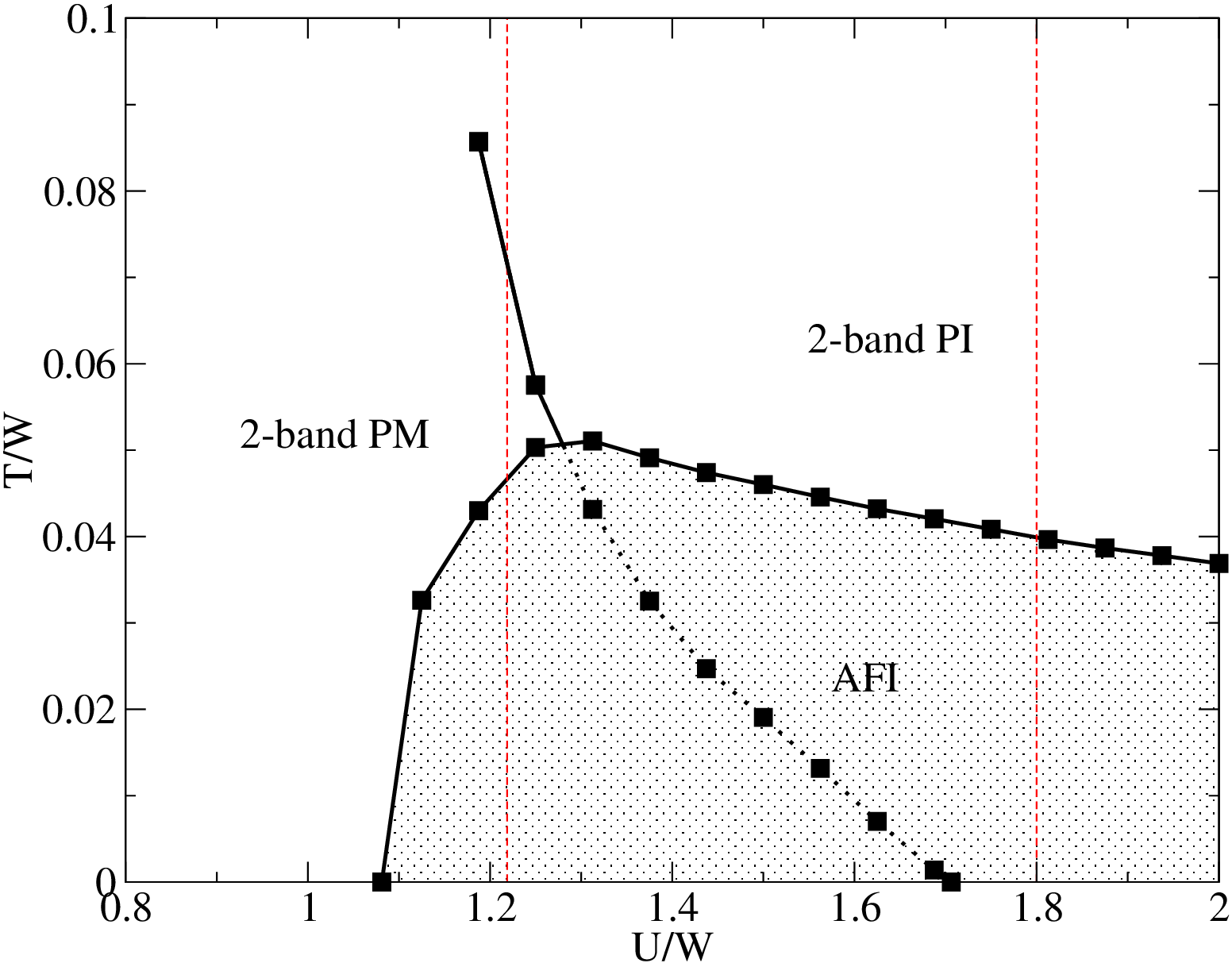}
\caption{(Color online) Finite temperature phase diagram as a function of U and T, for a fixed value of $\Delta = 0.025W$. The paramagnetic solution is continued also within the AFM domain (dotted line). The vertical red dashed line indicates the temperature cut represented in Fig.\ref{fig:phaseFT_R}.}
\label{fig:phaseFT}
\end{figure}

\begin{figure}[htb]
\includegraphics[scale=0.32]{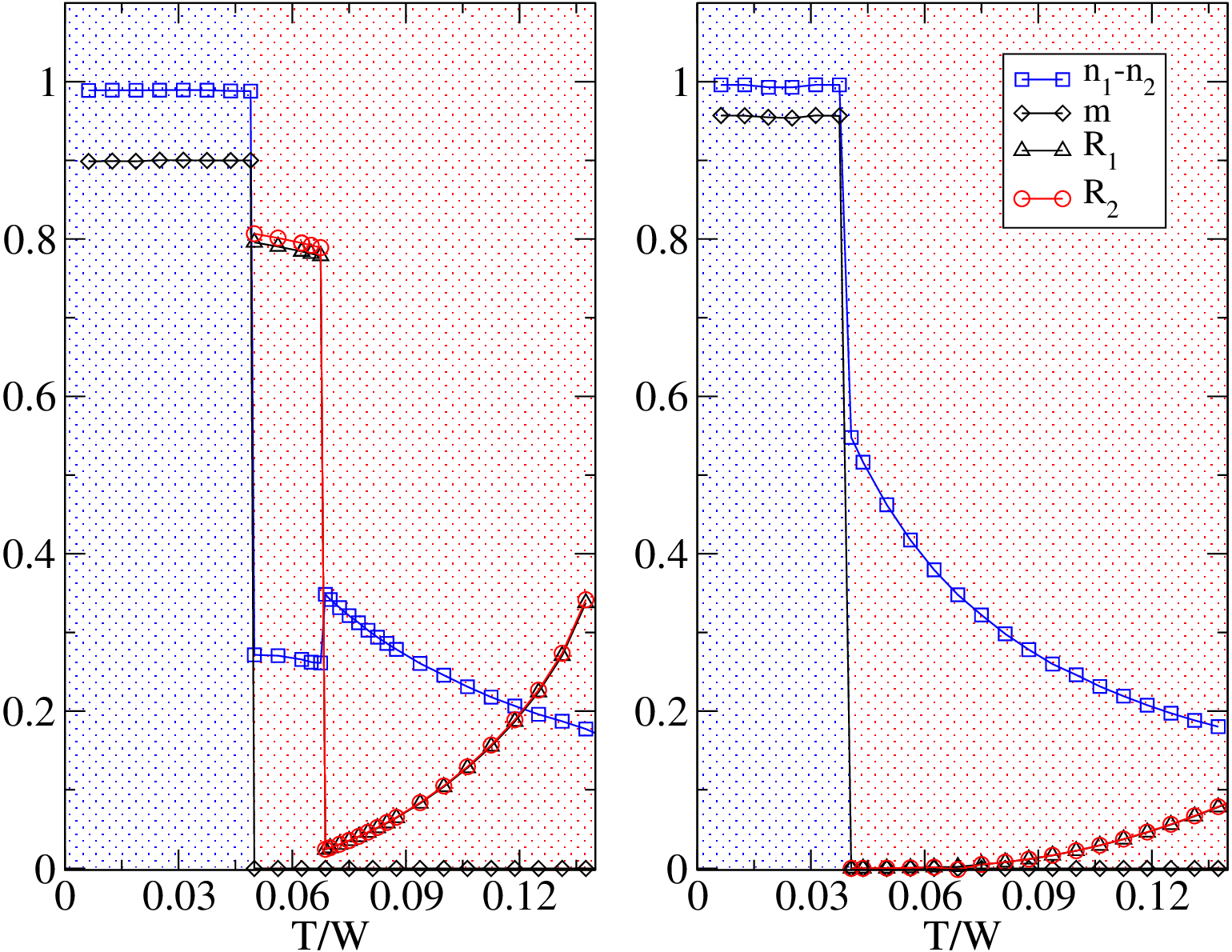}
\caption{(Color online) The blue (to the left in each panel) and red areas (to the right) indicate respectively the AFI phase and the paramagnetic phases as a function of temperature at fixed values of $U=1.22W$ (left panel) and $U=1.8W$ (right panel). At low temperatures the orbital polarization (blue squares) and the staggered magnetization (black squares) are practically equal to the zero temperature values and display a discontinuous jump at the AFI-paramagnetic transition. In the paramagnetic phase we show also the behavior of the renormalization factors whose jump indicates the PM-PI transition. }
\label{fig:phaseFT_R}
\end{figure}

When magnetism is allowed, at zero temperature and at large $U$ the ground state is antiferromagnetic. At finite temperature the system remains ordered up to the N\'eel temperature. 
In Fig. \ref{fig:phaseFT} we plot for $\Delta=0.025W$ the phase diagram, indicating by a dotted line the PM-PI transition that 
we have found in the paramagnetic sector. 
We note that the PM-PI transition line crosses the N\'eel temperature, 
roughly at $U\simeq 1.28W$, and extends above. For $U>1.28W$, the Gutzwiller variational approach is able to describe melting of the AFI into a 2-band PI, which we mentioned is not accessible by Hartree-Fock. For smaller values of $U$, the magnetic insulator turns into a 
2-band PM that eventually undergoes a Mott transition at higher temperatures. 
In Fig. \ref{fig:phaseFT_R} we show more in detail the behavior of the physical quantities across the different transitions; in the low temperature AFI  (blue area on the left) the orbital and magnetic polarizations are very weakly temperature dependent. Increasing $T$, the N\'eel order melts, the orbital polarization decreases (red areas on the right), and the model turns into a 2-band PM (left panel) or PI (right panel) depending on the value of $U$. In the former case, left panel of Fig. \ref{fig:phaseFT_R}, the 2-band PM is eventually driven to the PI state, 
transition that is signaled by the sudden vanishing of the renormalization factors and the jump of the orbital polarization $n_1-n_2$. In the right panel, instead, the AFI melts directly in the PI; the renormalization factor vanish at the transition and then smoothly increases from zero on raising $T$.

We  observe that the finite temperature phase diagram of Fig. \ref{fig:phaseFT} is not dissimilar to that of 
V$_2$O$_3$ as function 
of chemical/physical pressure.  Also the physical mechanism that controls the phase diagram, i.e. the correlation enhanced crystal-field splitting, is consistent with that proposed in Ref. \onlinecite{Georges&Andersen} for  
V$_2$O$_3$, though in our case the number of orbitals involved is two and not three. We also note the discontinuous increase of the orbital polarization across the PM to PI transition upon increasing temperature, see right panel in Fig. \ref{fig:phaseFT_R}, which is consistent 
with X-ray adsorption spectra  of V$_2$O$_3$,\cite{Park&Sawatzky,Marino-PRL} in which case the 
orbital polarization relates to the occupation of the $e^\pi_g$ orbitals with respect to the $a_{1g}$ one.

\begin{figure}[htb]
\includegraphics[scale=0.32]{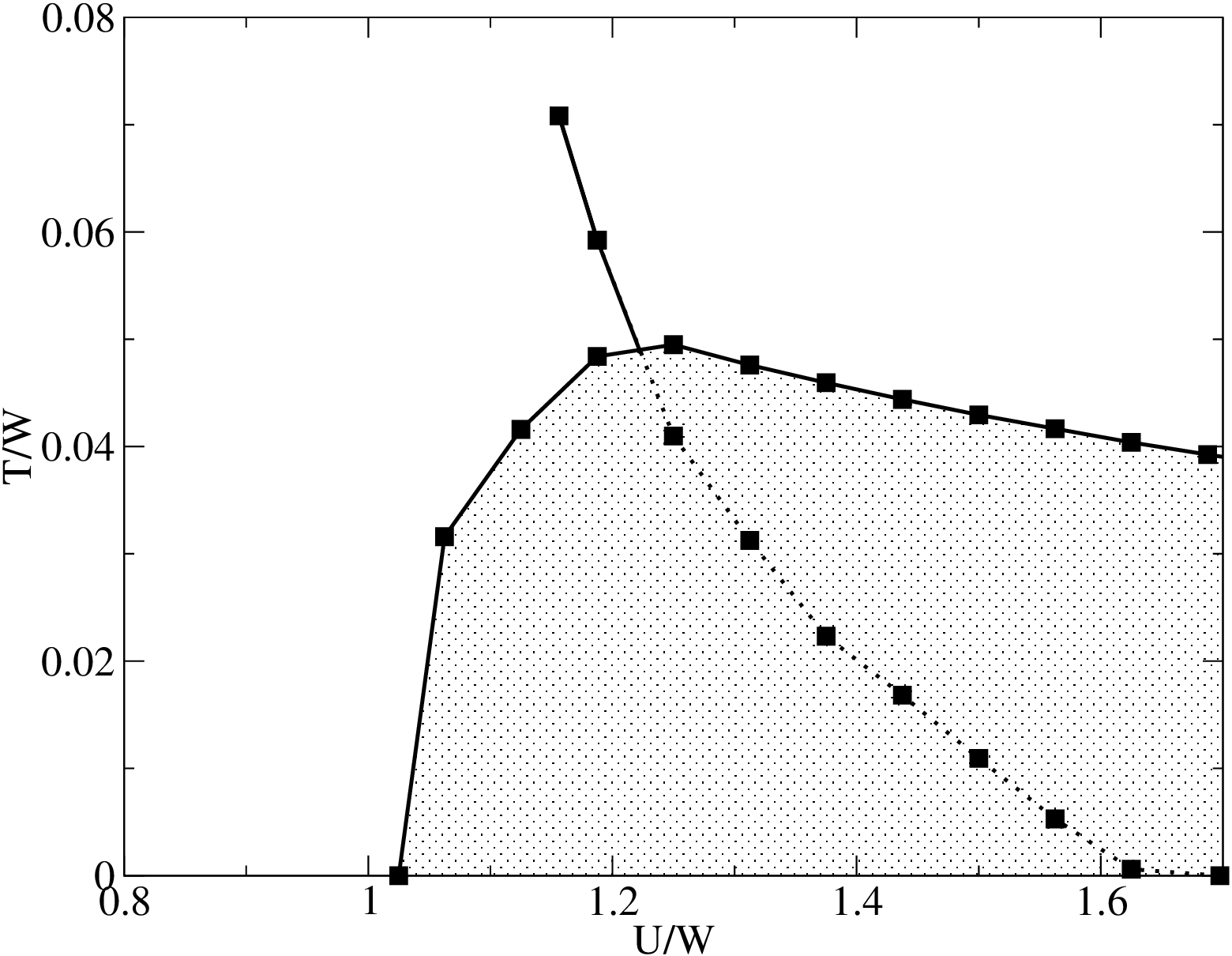}
\vspace{10pt}
\includegraphics[scale=0.32]{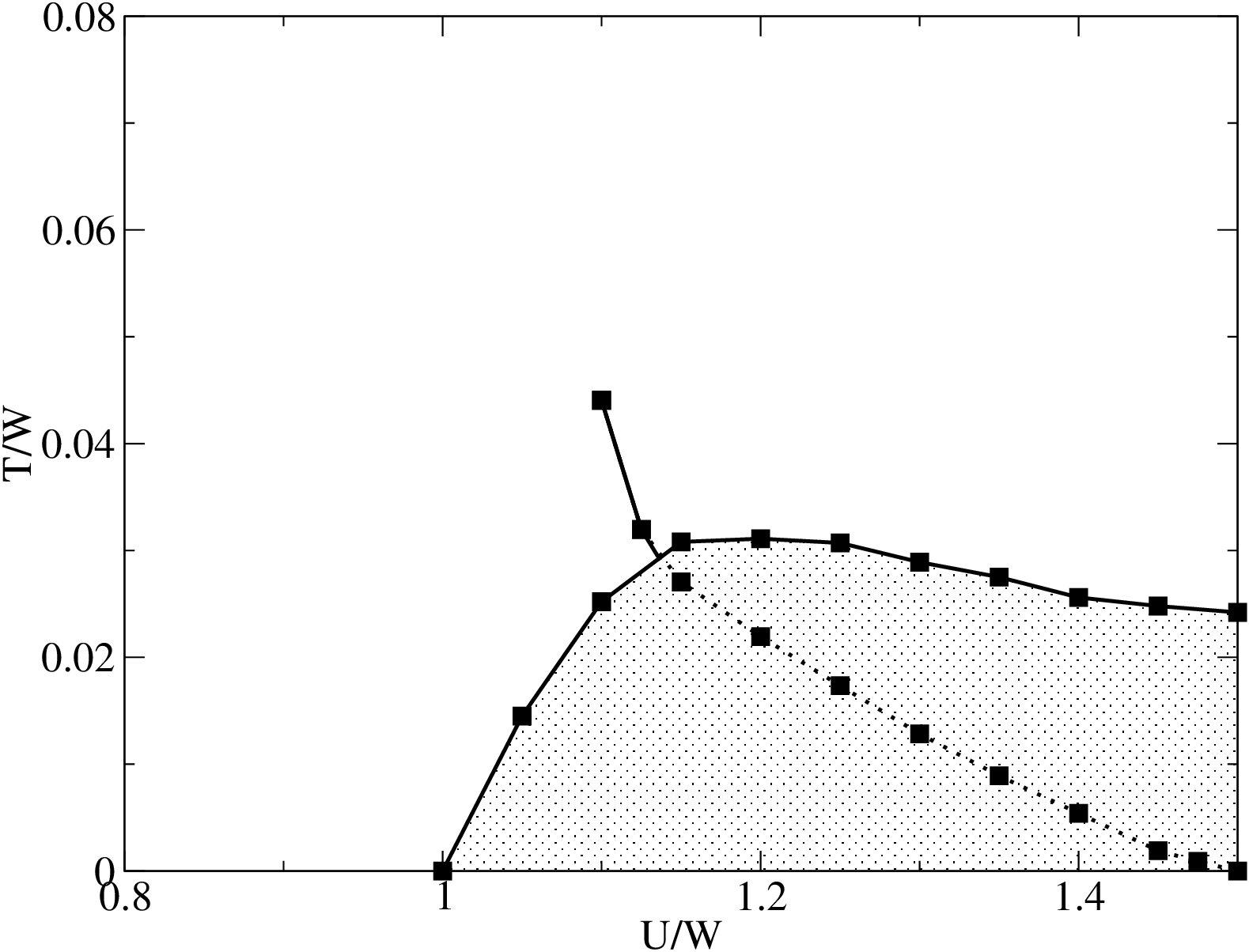}
\caption{Top panel: finite $T$ phase diagram within the Gutzwiller approximation at $\Delta=0.025W$, $t'=0$ and a semicircular DOS. Bottom panel: same as before but within DMFT, which is exact.}
\label{fig:phaseFT_compare}
\end{figure}

\section{Comparison with DMFT}
\label{cinque}

In this section, we compare the quality of the finite temperature 
Gutzwiller approximation with exact  DMFT results.  
In particular, we shall consider a simplified version of the model in Eq. \eqn{Ham} with vanishing inter-orbital hybridization, $t'=0$, and on a Bethe lattice with only nearest neighbor hopping, which leads to a non-interacting semicircular density of states of bandwidth $W$. 
We choose a Bethe lattice (a Cayley tree with coordination number $z\to\infty$) because in this case  DMFT is exact and, as previously discussed, the Gutzwiller approximation does provide a rigorous upper bound to the free energy, which therefore makes it possible to assess its accuracy with respect to exact results. 
The phase diagrams obtained by DMFT and by the Gutzwiller approximation in the $U$-$T$ space for $\Delta = 0.025W$ are shown in Fig. \ref{fig:phaseFT_compare}.

DMFT maps the lattice model onto an impurity model, which, in the
present calculation, is solved by means of exact
diagonalization\cite{caffarelkrauth} in the finite-temperature
implementation proposed in Ref.~\onlinecite{caponedemedicigeorges},
which is particularly accurate at the low temperatures that we
consider. Within the exact diagonalization approach, the bath is
approximated by a finite number, $N_b$, of energy levels. Here we take
$N_b = 10$ and $N_b=12$ , i.e. 5 and  6 bath levels for each physical
orbital. Only for $N_b=10$ we could include a number of states
sufficient to obtained converged results. Therefore data for $N_b =
12$ have only obtained for low temperatures, and used to prove that
the discretization error only leads to minor corrections to the phase diagram.
We consider both paramagnetic and antiferromagnetic solutions.  
As customary, we first determine the Mott transition line in the paramagnetic sector by comparing the free energies of the metallic and Mott insulating solutions. The transition is first-order at any finite temperature and ends in a finite-temperature critical point at $T \lesssim 0.05 W$.  If we allow for long-range antiferromagnetic order, at $T=0$ the system is metallic for $U \lesssim W$, and it turns into a single-band antiferromagnet for larger values of the interaction. The N\'eel temperature rapidly grows with $U$ and reaches a maximum around $U \simeq 1.2W$, above which it monotonically decreases. However, differently from the single-band case, the first-order Mott line is not completely covered by the antiferromagnetic dome. 

The DMFT phase diagram is thus very similar to that obtained by the finite-temperature Gutzwiller approach, qualitatively and to same extent also quantitatively, see Fig. \ref{fig:phaseFT_compare}.  As common with the Gutzwiller approximation, the $T=0$ Mott transition in the paramagnetic sector occurs at larger $U/W\simeq 1.7$ than 
the exact DMFT value $U/W\simeq 1.5$. In addition, the Gutzwiller wavefunction seems to overestimate antiferromagnetism, which occupies a larger region in the phase diagram. However, quite remarkably, the critical endpoints of the PM-PI Mott transition do not differ much, 
$U/W\simeq 1.17$ and $T/W\simeq 0.07$ in the Gutzwiller calculation, while $U/W \simeq 1.15$ and $T/W\simeq 0.05$ in DMFT.

\section{Conclusions}
\label{sei}

Using some rigorous trace inequalities, we have derived an upper-bound estimate of the free energy 
of an interacting-electron Hamiltonian for variational density
matrices of  Gutzwiller and Jastrow type.
We have then exploited this result to extend to finite temperature the
conventional Gutzwiller approximation, which in turn becomes an exact variational approach in lattices with infinite coordination number. 

We have applied this technique to calculate the finite-temperature
phase diagram of a two-band model that we believe captures
qualitatively well the physics of vanadium sesquioxide, V$_2$O$_3$. In
spite of being extremely simplified with respect to a complete
description of V$_2$O$_3$, the model has a very similar phase diagram comprising a
low-temperature antiferromagnetic insulating dome and high-temperature
paramagnetic metal as well as Mott insulating phases separated by a
first order line with a second-order critical endpoint. We have tested
the accuracy of our finite temperature Gutzwiller approximation
comparing the phase diagram of the model on a Bethe lattice with the
exact one obtained by DMFT. The agreement is qualitatively very
satisfying and partly also quantitatively.  We believe therefore that this simple variational technique is 
very promising to attach correlated electron systems at finite
temperature, and could be used whenever more reliable tools, like
DMFT, become numerically too demanding.

\acknowledgments
We thank Nicola Lanat\`a and Giovanni Borghi for very useful discussions 
and helpful suggestions. 
This work has been supported by the European Union, Seventh Framework
Programme, under the project GO FAST, Grant Agreement no. 280555 and
the European Research Council Starting Grant SUPERBAD, Grant Agreement no. 240524.


\end{document}